\documentclass{article}
\usepackage[utf8]{inputenc}
\usepackage[inline]{enumitem}
\usepackage{graphicx}
\usepackage{float}
\usepackage{geometry}
\usepackage{amsmath}
\usepackage{amssymb}
\usepackage{color}
\usepackage[hidelinks]{hyperref}
\usepackage{microtype}
\usepackage{authblk}
\usepackage{subcaption}
\usepackage{titlesec}

\titlespacing*{\section}{0pt}{6pt plus 2pt minus 2pt}{2pt plus 1pt minus 1pt}
\titlespacing*{\paragraph}{0pt}{4pt plus 2pt minus 1pt}{1em}

\newcommand{\EBL}{\mathrm{EBL}}
\newcommand{\E}{\mathbb{E}}
\newcommand{\SE}{\mathrm{SE}}
\newcommand{\bp}{\,\mathrm{bp}}

\title{\vspace{-0.5in}Neutral Is Not Free: Evaluating Downside Risk in Neutral Launches}
\author{Pablo Alcain}
\author{Jason Kang}
\author{Mia Garrard}
\author{Marine Veits}
\author{Houssam Nassif}
\author{Abbas Zaidi}
\affil{Meta Platforms, Inc.}
\date{}

\begin{document}

\setlength{\abovedisplayskip}{4pt}
\setlength{\belowdisplayskip}{4pt}
\setlength{\abovedisplayshortskip}{0pt}
\setlength{\belowdisplayshortskip}{4pt}

\maketitle
\vspace{-0.2in}

\section{Introduction}
\label{sec:intro}

Online randomized controlled trials (A/B tests) are the standard for quantifying the impact of product decisions prior to release~\cite{kohavi2020trustworthy}. However, as the resource costs associated with ideation and implementation continue to fall, a growing share of experiments no longer attempt to explore improvements along top-line decision metrics~\cite{gupta2019top}. Instead, these \emph{neutral launches}—which include codebase migrations, dependency upgrades, infrastructure reliability work, and resource reduction changes—focus on maintaining the status quo while capturing underlying operational benefits.

Despite their unique objectives, most experimentation platforms still evaluate neutral launches using a confidence interval (CI) overlap rule: an experiment is deemed safe to launch if its confidence interval contains zero (formally, $|\hat\theta| \le z_{\alpha/2}\,\SE(\hat\theta)$). This approach is conceptually inadequate, as it conflates the absence of evidence with evidence of harmlessness~\cite{altman1995absence,wasserstein2016asa}. In practice, this theoretical flaw manifests in two major operational issues. First, the rule is perverse with respect to information: it is dangerously permissive when data is scarce (since a large standard error guarantees overlap), and excessively restrictive when data is abundant (where a tiny standard error classifies a neutral treatment as a negative one). Second, it yields a binary outcome, offering decision-makers no quantifiable information about the expected magnitude of harm.

To find a more robust framework, we look to early-stage dose escalation in Phase I clinical trials, which face similar challenges. Rather than proving therapeutic superiority, these trials seek to establish safety and identify a Maximum Tolerated Dose (MTD) under an acceptable toxicity limit. Instead of relying on CI overlap assessments, modern trial designs—such as the Continual Reassessment Method (CRM)~\cite{oquigley1990continual} and Escalation with Overdose Control (EWOC)~\cite{babb1998cancer}—utilize Bayesian downside-risk models to explicitly bound expected harm.

In the realm of A/B testing, the most rigorous frequentist alternative is the binary non-inferiority test~\cite{blackwelder1982proving,lakens2017equivalence}. Unfortunately, because acceptable non-inferiority margins are typically much smaller than the effect sizes required in superiority tests, achieving the necessary sample size is notoriously difficult. Consequently, practitioners frequently abandon non-inferiority testing in favor of the flawed CI-overlap rule, occasionally patching it with ad-hoc standard error requirements. Others rely on posterior regression probability, which mitigates some issues but still fails to quantify the magnitude of the effect.

We address these challenges by introducing the Expected Bayesian Loss (EBL)—the posterior expectation of a downside-only loss~\cite{berger1985statistical,fishburn1977mean}—as a continuous evaluation metric for neutral launches. By deriving a closed-form solution for EBL that leverages existing frequentist reporting, we characterize its distinct mathematical advantages over standard alternatives. We demonstrate EBL's empirical performance on an expert-reviewed cohort of experiments, proving its efficacy as a tunable policy boundary that aligns statistical guardrails with institutional risk exposure.

The remainder of this paper is organized as follows. Section~\ref{sec:method} formally introduces the EBL framework and derives its closed-form representation. Section~\ref{sec:frequentist} connects these Bayesian expectations to frequentist properties, sampling behaviors, and the principles of Calibrated Bayes. Section~\ref{sec:results} empirically validates the metric against expert committee decisions using a historical experimentation cohort. Section~\ref{sec:governance} outlines practical considerations for the operational governance of EBL guardrails. Finally, Section~\ref{sec:conclusion} concludes the paper and sets up future directions, including the integration of hierarchical shrinkage priors.

\section{EBL Policy Construction}
\label{sec:method}

\paragraph{Policy Formulation.}
To construct a rigorous neutral-launch policy, we compare the expected harm of a treatment effect $\theta$ against two predefined thresholds: a strict \emph{blocking} threshold $\tau_B$ and a lower \emph{reviewing} threshold $\tau_R < \tau_B$. Given experiment data $D$, the operational decision rule is:
\begin{equation*}
    \EBL(D) > \tau_B \Rightarrow \textbf{block} , \quad
    \textit{else } \EBL(D) \ge \tau_R \Rightarrow \textbf{review} , \quad
    \textit{else } \textbf{launch}.
\end{equation*}
Formally, we define the loss of launching as $L(\text{launch},\theta)=(-\theta)_+$, where $(-\theta)_+ = \max(0, -\theta)$. Its posterior expectation naturally factors into two components—the probability of harm and the severity of that harm:
\begin{equation*}
    \EBL(D)=\E\!\left[(-\theta)_+\mid D\right]
    =\underbrace{P(\theta<0\mid D)}_{\text{chance of harm}}\;\cdot\;\underbrace{\left[-\E(\theta\mid\theta<0,D)\right].}_{\text{severity given harm}}
\end{equation*}
This factorization yields a non-negative quantity expressed in the original units of $\theta$. Should the platform tolerate degradations down to a specific margin $-\Delta$, the shifted loss $(-\theta-\Delta)_+$ can be seamlessly substituted to evaluate only the harm exceeding that tolerance.

\paragraph{Closed Form from a Reported Interval.}
Translating this Bayesian formulation into practice often requires complex joint modeling. However, by relying on the Central Limit Theorem and applying a formal, non-informative Jeffreys prior ($p(\mu) \propto 1$)~\cite{jeffreys1946invariant}, we can reliably approximate the posterior as $\mu\sim\mathcal{N}(m,s^2)$, where $m=\hat\theta$ and $s=\SE(\hat\theta)$. This simplification allows us to extract Bayesian expectations directly from existing frequentist analysis infrastructure. Substituting $z=(\theta-m)/s$ yields a closed-form solution:
\begin{equation*}
\EBL(D)
=-\int_{-\infty}^{-m/s}(m+sz)\phi(z)\,dz
= s\,\phi\!\left(\frac{m}{s}\right)-m\,\Phi\!\left(-\frac{m}{s}\right).
\end{equation*}
Thus, the expected loss can be entirely reconstructed from standard frequentist point estimates and confidence intervals.

\paragraph{Probability Versus Severity.}
To understand how EBL improves upon simple posterior regression probability, we can standardize the estimate by defining $u=m/s$. Under this parameterization, the two risk scores cleanly separate:
\begin{equation*}
P(\theta<0\mid D)=\Phi(-u) \quad \text{and} \quad 
\EBL(D)=s\,g(u), \quad \text{with} \quad
g(u)=\phi(u)-u\,\Phi(-u).
\end{equation*}
Notice that while regression probability is solely a function of $u$, EBL depends on both $s$ and $u$. Consider two experiments $i$ and $j$ exhibiting the same probability of regression ($u_i = u_j$). The EBL framework will penalize the experiment with the higher variance ($s$), explicitly identifying it as the riskier launch. By scaling expected loss to account for empirical noise, EBL successfully inverts the CI-overlap rule's fatal pathology: the metric becomes \emph{harder} to pass as the estimate grows noisier. Regression probability and EBL only yield identical orderings when standard errors are exactly equal across experiments.

\paragraph{Evaluating the Launch Decision.}
A statistical guardrail must be operationally defensible and comparable across multiple experiments. The rule ``block if $\EBL>\tau_B$'' operates as a formal, institutional assertion that a given class of technical launches is worth up to $\tau_B$ in metric degradation. Setting $\tau_B$ empowers stakeholders to explicitly define their risk appetite for neutral launches and track it consistently across time. The metric is inherently transparent: it quantifies exactly how much metric value the platform expects to lose on average by enacting the treatment, independent of any external operational gains.

\section{Frequentist Properties and Sampling Characteristics}
\label{sec:frequentist}

To bridge the gap for frequentist practitioners, we now demonstrate how EBL—despite being a Bayesian expectation—functions as a deterministic transformation of frequentist statistics and possesses definable long-run sampling behaviors.

\paragraph{Deriving the EBL Cumulative Distribution Function (CDF).}
When treating the frequentist point estimate $m$ as a random variable $M$ over repeated sampling, it follows $M \sim \mathcal{N}(\theta, s^2)$. By defining the standardized test statistic $Z = M/s$, which distributes as $Z \sim \mathcal{N}(\theta/s, 1)$, we can express the EBL as a random variable $\EBL = s \cdot g(Z)$. Because $g(u)$ is strictly monotonically decreasing, it is mathematically invertible. Consequently, the exact probability of the expected loss falling below a threshold $e$ over repeated sampling is:
\begin{equation*}
P(\EBL \le e) = 1 - \Phi\left(g^{-1}\left(\frac{e}{s}\right) - \frac{\theta}{s}\right).
\end{equation*}

\paragraph{Mapping to Type I Error Rates.}
In a classical non-inferiority framework, fixing a block threshold ($\tau_B$) maps directly to frequentist error rates depending on the defined null hypothesis. We consider two primary operational errors:

\textit{Error 1: Falsely Blocking a Neutral Launch.} If a launch is perfectly neutral ($\theta = 0$), the probability that empirical noise artificially pushes the EBL over $\tau_B$ is:
\begin{equation*}
\alpha_{\text{block}} = P(\EBL > \tau_B \mid \theta = 0) = \Phi\left(g^{-1}\left(\frac{\tau_B}{s}\right)\right).
\end{equation*}

\textit{Error 2: Falsely Launching a Degradation.} Conversely, if the true effect rests exactly at the maximum acceptable harm ($\theta = -\Delta$), the probability the EBL stays below $\tau_B$ is:
\begin{equation*}
\alpha_{\text{launch}} = P(\EBL \le \tau_B \mid \theta = -\Delta) = 1 - \Phi\left(g^{-1}\left(\frac{\tau_B}{s}\right) + \frac{\Delta}{s}\right).
\end{equation*}
Because $g^{-1}$ has no closed-form solution, these specific error boundaries must be evaluated numerically for any given variance $s^2$.

\paragraph{Informative Priors and Calibrated Bayes.}
While the closed-form EBL relies on a non-informative Jeffreys prior~\cite{jeffreys1946invariant} to perfectly match frequentist reporting, incorporating an informative conjugate Gaussian prior, $\mu \sim \mathcal{N}(\mu_0, \sigma_0^2)$, fundamentally alters these sampling characteristics. Under a precision-weighted posterior, the expected value of the standardized test statistic is pulled toward the prior mean ($\mu_0$), and its overall sampling variance is strictly reduced by the shrinkage factor $B = \sigma_0^2 / (\sigma_0^2 + s^2)$. 

To ensure these altered error rates remain operationally defensible, we can adopt the principles of Calibrated Bayes~\cite{zhou2023bayesian}. Rather than selecting subjective prior parameters, a calibrated approach estimates $\mu_0$ and $\sigma_0^2$ directly from the empirical distribution of historical neutral launches on the platform. By aligning the prior with the true data-generating process, the shrinkage applied to the EBL ceases to be an arbitrary subjective bias and instead becomes an optimal frequentist adjustment that minimizes the long-run Mean Squared Error (MSE) across the platform's portfolio. Consequently, the frequentist operating characteristics shift from conditional guarantees to marginal ones. Instead of evaluating the false-block rate ($\alpha_{\text{block}}$) conditionally on a single, fixed true effect (e.g., $\theta = 0$), a Calibrated Bayesian framework evaluates the expected error rate integrated over the empirical prior distribution of all true effects. This ensures that the EBL thresholds yield valid long-run frequentist coverage.

\section{Empirical Findings and Discussion}
\label{sec:results}

\begin{figure}[th]
\centering
  \begin{subfigure}[t]{0.48\textwidth}
  \centering
  \includegraphics[width=\linewidth]{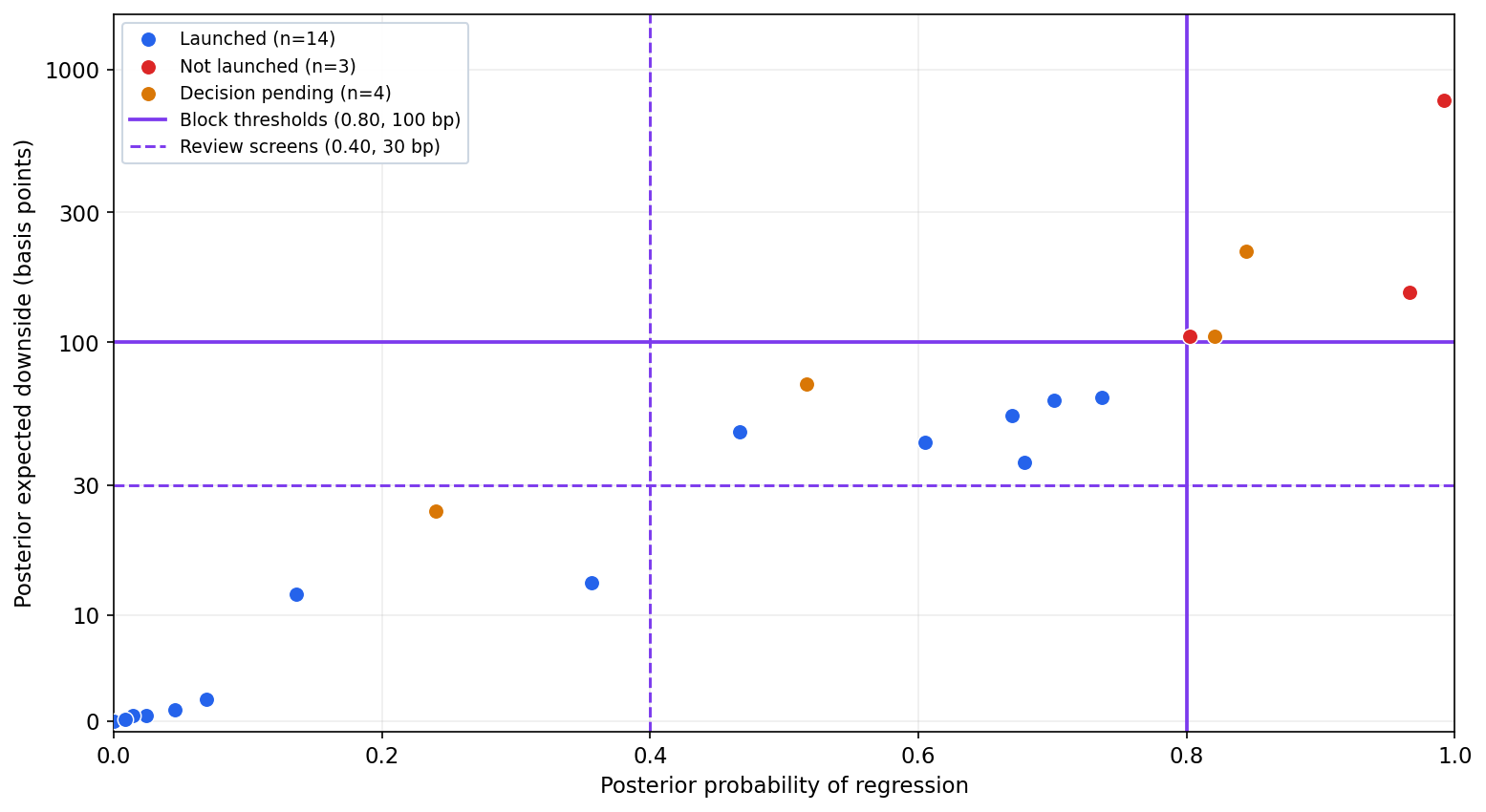}
  \caption{Historical decisions for EBL and regression probability.}
  \label{fig:historical-decisions}
  \end{subfigure}\hfill
  \begin{subfigure}[t]{0.48\textwidth}
  \centering
  \includegraphics[width=\linewidth]{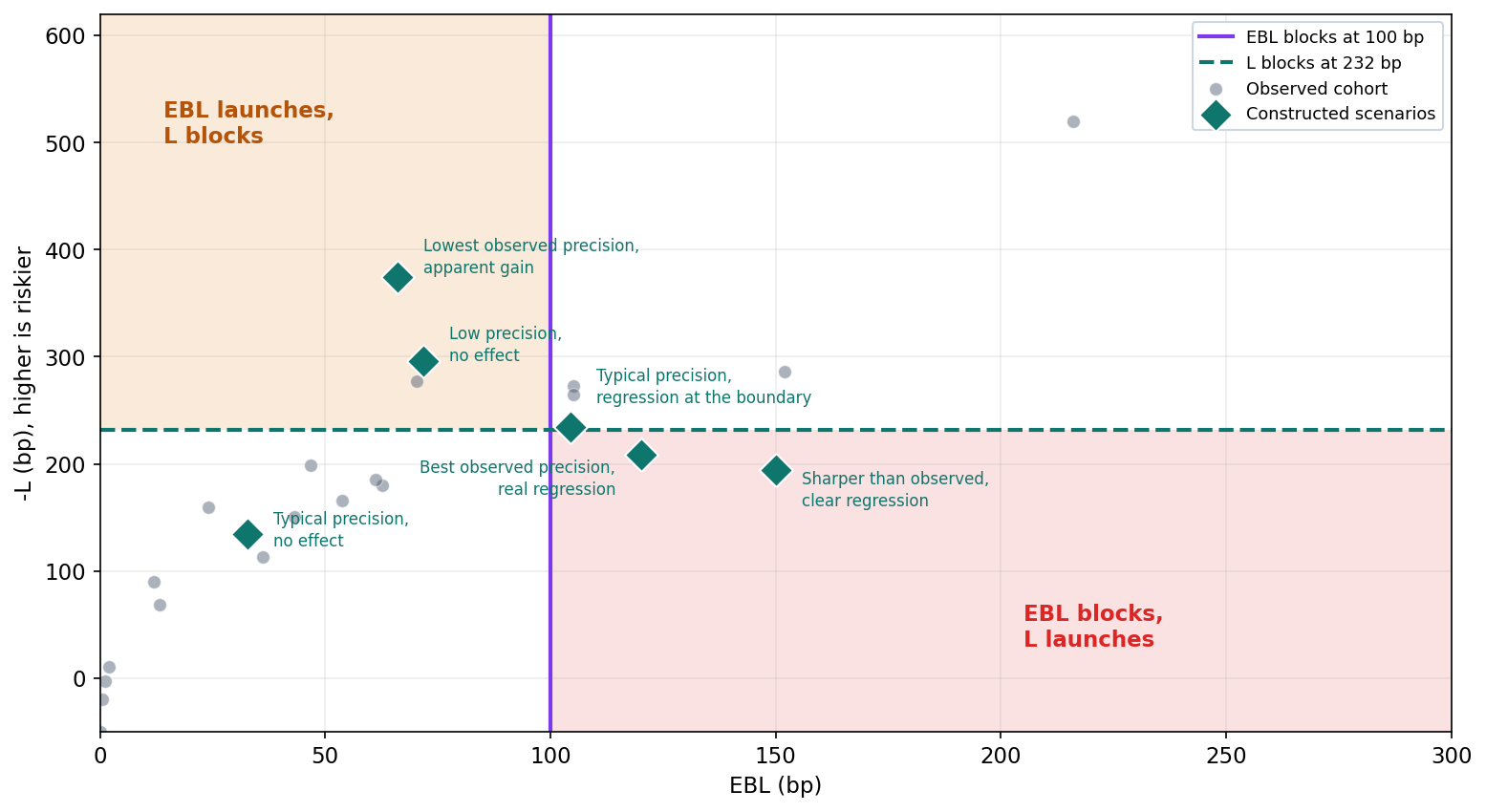}
  \caption{Comparison between EBL and Lower bound.}
  \label{fig:comparison}
  \end{subfigure}
  \caption{(a) Posterior regression probability and EBL across the anonymized portfolio, including the four pending cases. Solid lines mark the $80\%$ and $100$-basis-point block thresholds; dashed lines mark the $40\%$ and $30$-basis-point review screens. (b) Comparison between EBL and the Lower bound policy in the anonymized portfolio, highlighting examples where the two methods differ.}
  \label{fig:decisions}
\end{figure}

To empirically validate the framework~\cite{Malmir2026ImpressionShare}, we analyzed a historical cohort of 17 experiments exhibiting highly similar precision $s$. Because neutral launches lack a definitive objective ground truth, we utilized recorded expert committee decisions as labels (14 launched, 3 blocked), alongside four pending cases. Here, $\theta$ represents the relative lift in the metric of interest, and we report EBL such that $100\bp=1\%$ relative degradation. We successfully replicated the experts' decisions by setting $\tau_B=100\bp$, a threshold that matches a regression probability of $\sim80\%$ given the cohort's precision. Following further review with the experts, we also established lower review screens at $30\bp$ and $40\%$ (see Figure~\ref{fig:historical-decisions}). To prevent heavily imbalanced data from artificially inflating our performance metrics, we evaluated policies using leave-one-out (LOO) per-class agreement. 

Although both EBL and regression probability classified this specific cohort equally well, they are not mathematically interchangeable. As derived in Section~\ref{sec:method}, the two scores only rank experiments identically when posterior precision ($s$) is near-constant across the cohort. Within our dataset, $s$ varied by a factor of 1.95, while $g(u)$ varied by a much larger factor of $\sim26$. This indicates that the observed agreement between the two methods is merely an artifact of this specific portfolio's narrow variance spread. 

EBL succeeds precisely where regression probability fails: guarding against highly uncertain results. An experiment with massive variance might easily pass an $80\%$ regression probability threshold, but it will correctly fail the $100\bp$ EBL threshold because EBL explicitly scales risk against empirical noise. The divergence condition ($s_i/s_j > g(u_j)/g(u_i)$) dictates exactly which future, high-variance experiments will break this mathematical tie, highlighting exactly where probability alone leaves a platform dangerously unprotected.

The primary test of the proposed guardrail is its ability to reliably replicate nuanced expert decisions better than platform defaults. In-sample, an EBL threshold of $100\bp$ and a regression-probability threshold of $80\%$ successfully flagged all three no-launch decisions and accepted all 14 launches. Evaluated leave-one-out, both frameworks scored $14/14$ on launches and $2/3$ on no-launches. We also compared this region split against an alternative provided by standard frequentist metrics: the left-boundary of the confidence interval. While this lower-bound policy can be artificially tuned to reproduce the block boundary in this low-variance dataset, it completely fails to highlight a review region that agrees with experts' views, achieving an accuracy of $<80\%$ compared to the $>95\%$ accuracy provided by EBL. To further illustrate this structural flaw, Figure~\ref{fig:comparison} compares the \emph{block} boundaries created by EBL and the Lower bound policy (labeled as L). If we allow for varying $s$, the two policies completely decouple, and the Lower bound policy begins allowing launches that are clear regressions and should be explicitly blocked.

Ultimately, the true value of EBL lies not in raw classification accuracy, but in its ability to holistically evaluate the decision. Platform reviewers can explicitly weigh the expected continuous loss against a migration's tangible engineering benefits, progressively tuning the boundaries in accordance with institutional risk appetite. Conversely, regression probability reports only a scalar chance of a negative sign, offering no actionable magnitude to trade against external operational benefits.

\section{Operational Governance of EBL Guardrails}
\label{sec:governance}

While the mathematical framework successfully describes expected-loss guardrails, widespread operationalization should include centralized review boards to enforce these properties systematically across an experimentation platform. 

\paragraph{Setting Institutional Risk Appetite.}
A centralized governance board can define the acceptable degradation limit ($\Delta$) and the blocking threshold ($\tau_B$). By fixing $\tau_B$, the board creates a known, computable trade-off that forces the organization to formally declare exactly how much expected metric loss it is willing to absorb in exchange for engineering velocity.

\paragraph{Enforcing Experimental Rigor.}
As established in Section~\ref{sec:frequentist}, the false-block rate ($\alpha_{\text{block}}$) strictly increases as the standard error ($s$) increases. This built-in mathematical feature acts as an automated governance mechanism: it actively penalizes empirical noise, thereby preventing engineering teams from submitting underpowered, high-variance experiments to bypass guardrails. If a team submits a noisy test, the calculated EBL artificially inflates, immediately triggering an automatic block and forcing the team to collect more data.

\paragraph{Governing Multi-Metric Impact.}
Just as clinical trials require complex joint modeling to account for multiple competing toxicities~\cite{thall1998strategy}, EBL should not be naively aggregated across multiple platform guardrails without explicit conversion weights. To manage complex platforms, we propose two distinct governance models:
\begin{enumerate*}[label=(\roman*)]
\item \textbf{The Min-Max Policy:} The launch is strictly governed based on the single worst metric's EBL. This aggressive approach is most suitable for non-negotiable health metrics, such as system latency or server error rates.
\item \textbf{The Weighted Loss Policy:} The launch is regulated based on a weighted sum of EBLs across affected metrics. The governance board can establish explicit, well-calibrated conversion weights between fundamentally disparate metrics.
\end{enumerate*}

\paragraph{Protecting Against Misspecification.}
Because EBL operates as a left-tail functional, it is inherently more sensitive to posterior misspecification (such as heavy-tailed distributions) than simple regression probability. Therefore, calibration heavily depends on the expected true distribution of effects. Governance boards should mandate a standardized calibration process based on the platform's historical effect distributions to prevent practitioners from gaming the output via optimistic variance reporting. Ultimately, what EBL resolves today is the ambiguity of the underlying operational metric: neutral-launch review stops emitting a vague statistical significance label and instead emits a precise, actionable value.

\section{Conclusion}
\label{sec:conclusion}

The Expected Bayesian Loss (EBL) framework provides a robust, continuous evaluation metric for neutral launches, successfully resolving the severe theoretical and operational limitations of the standard frequentist confidence interval overlap rule. By defining risk as a function of both the probability of metric regression and its expected severity, EBL equips practitioners with a mathematically sound, operationally transparent threshold that directly scales risk against empirical noise. Utilizing a non-informative Jeffreys prior, we have shown that this Bayesian expectation can be extracted seamlessly from existing frequentist reporting infrastructure, allowing centralized governance boards to codify explicit risk appetites without overhauling platform reporting.

While this paper relies on a non-informative prior to maintain a direct mapping to frequentist outputs, adopting an informative conjugate Gaussian prior fundamentally alters the metric's operating characteristics. Under a precision-weighted posterior, the expected value of the standardized test statistic is pulled toward the prior mean, and its overall sampling variance is strictly reduced. Operationally, this sacrifices classical frequentist unbiasedness in favor of stability; it systematically reduces the probability of falsely blocking a noisy neutral launch ($\alpha_{\text{block}}$), while demanding much stronger empirical evidence of degradation to correctly block a harmful launch ($\alpha_{\text{launch}}$).

Building on the principles of Calibrated Bayes discussed in Section~\ref{sec:frequentist}, future work will explore replacing simple static priors with a more sophisticated hierarchical structure, specifically the Bayesian Hybrid Shrinkage (BHS) approach proposed by Mudd et al.\ (2025)~\cite{mudd2025breaking}. BHS directly actualizes the calibrated objective by utilizing an inverse-Gamma hyperprior to empirically learn and regulate experiment-specific local shrinkage across the testing portfolio. Because this hierarchical structure dynamically aligns the prior with the true data-generating process, it ensures valid long-run marginal coverage without requiring arbitrary, subjective parameter tuning. 

For centralized governance boards, this translates directly to enhanced operational robustness. First, BHS maintains high interval coverage even under heavy-tailed misspecifications, fulfilling the Calibrated Bayes mandate of minimizing long-run error across the platform while significantly reducing the administrative burden of manual calibration. Second, its adaptive local shrinkage provides a dynamic penalty for noisy submissions, aggressively correcting the estimates of underpowered experiments that attempt to bypass review thresholds. Applying this hybrid shrinkage framework allows the platform to proactively counteract the "Safe" Winner's Curse—a pervasive selection bias where teams repeatedly iterate and launch variants that pass guardrails purely due to favorable sampling variance. BHS explicitly corrects for this selection-induced bias, protecting the platform from cumulative metric erosion over time. Furthermore, because the BHS estimator ensures that shrinkage vanishes for extreme observations, governance boards can remain confident that massive, unambiguous metric degradations will never be artificially shrunk into the "safe" zone. Catastrophic errors will still easily breach the EBL threshold and trigger an automatic block, paving the way for even more rigorously calibrated platform protections.

\bibliographystyle{unsrt}
\bibliography{refs}

\end{document}